\documentclass[prx,twocolumn,
amsfonts,superscriptaddress,showpacs,longbibliography]{revtex4-2} 
\usepackage{natbib}
\usepackage{listings}
\usepackage{color}

\usepackage[utf8]{inputenc}
\usepackage{comment}

\definecolor{dkgreen}{rgb}{0,0.6,0}
\definecolor{gray}{rgb}{0.5,0.5,0.5}
\definecolor{mauve}{rgb}{0.58,0,0.82}

\usepackage[normalem]{ulem}

\usepackage{cancel}
\usepackage{amsmath,amssymb,mathrsfs}
\usepackage{latexsym}
\usepackage{graphicx} 
\usepackage{grffile}
\usepackage{epstopdf}
\usepackage{graphicx,epstopdf,color}
\usepackage{amsfonts}
\usepackage{amsmath,amssymb,mathrsfs}
\usepackage{hyperref}
\usepackage[dvipsnames]{xcolor}
\usepackage{url}
\usepackage{soul}
\usepackage{cancel}
\usepackage{appendix}
\usepackage{cleveref}
\crefname{equation}{Eq.}{Eqs.}
\Crefname{equation}{Equation}{Equations}
\crefname{figure}{Fig.}{Figs.}
\Crefname{figure}{Figure}{Figures}
\crefname{section}{Sec.}{Secs.}
\crefname{subsection}{Subsec.}{Subsecs.}
\Crefname{section}{Section}{Sections}
\crefname{appendix}{Appendix}{Apps.}
\Crefname{appendix}{Appendix}{Apps.}
\crefname{paragraph}{Sec.}{Secs.}
\crefname{table}{Table}{Tables}

\usepackage{bm}
\newcommand{\textalert}[1]{}

\newcommand{\ket}[1]{\left|#1\right\rangle}
\newcommand{\bra}[1]{\left\langle#1\right|}

\newcommand{\eps}{\varepsilon}

\newcommand{\half}{\frac{1}{2}}

\def\ie{i.e.\ }
\def\eg{e.g.\ }

\usepackage{xr}

\graphicspath{{nb_fig/}}

\usepackage[caption=false]{subfig}
\usepackage{epstopdf}

\newcommand{\hn}{\hat{n}}
\newcommand{\hphi}{\hat{\varphi}}

\renewcommand{\Im}{\text{Im}}
\renewcommand{\Re}{\text{Re}}

\newcommand{\RN}[1]{
\textup{\uppercase\expandafter{\romannumeral#1}}
}

\makeatletter
\def\@fnsymbol#1{\ensuremath{\ifcase#1\or * \or \mathsection\or \mathparagraph\or \|\or **\or \else\@ctrerr\fi}}
\makeatother

\begin{document}
\title{Strongly driven transmon as an incoherent noise source}

\author{Linda Greggio}
\thanks{linda.greggio@inria.fr}
\affiliation{Laboratoire de Physique de l’Ecole Normale Supérieure, Mines Paris, Inria,
CNRS, ENS-PSL, Sorbonne Université, PSL Research University, Paris, France}
\author{Rémi Robin}
\affiliation{Laboratoire de Physique de l’Ecole Normale Supérieure, Mines Paris, Inria,
CNRS, ENS-PSL, Sorbonne Université, PSL Research University, Paris, France}
\author{Mazyar Mirrahimi}
\affiliation{Laboratoire de Physique de l’Ecole Normale Supérieure, Mines Paris, Inria,
CNRS, ENS-PSL, Sorbonne Université, PSL Research University, Paris, France}
\author{Alexandru Petrescu}
\thanks{alexandru.petrescu@minesparis.psl.eu}
\affiliation{Laboratoire de Physique de l’Ecole Normale Supérieure, Mines Paris, Inria,
CNRS, ENS-PSL, Sorbonne Université, PSL Research University, Paris, France}

\date{\today}
\begin{abstract}
Under strong drives, which are becoming necessary for fast high-fidelity operations, transmons can be structurally unstable. Due to chaotic effects, the computational manifold is no longer well separated from the remainder of the spectrum, which correlates with enhanced offset-charge sensitivity and destructive effects in readout. We show here that these detrimental effects can further propagate to other degrees of freedom, for example to neighboring qubits in a multi-qubit system. Specifically, a coherently driven transmon can act as a source of incoherent noise to another circuit element coupled to it. By using a full quantum model and a semiclassical analysis, we perform the noise spectroscopy of the driven transmon coupled to a spectator two-level system (TLS), and we show that, in a certain limit, the interaction with the driven transmon can be modeled as a stochastic diffusive process driving the TLS. 
\end{abstract}
\maketitle

\section{Introduction}
Superconducting circuit quantum electrodynamics has become one of the most promising platforms for quantum information processing thanks to advances in state preparation, readout, and control \cite{blais_et_al_2021,krantz_et_al_2019}. The speed and fidelity of such operations are limited, however, by effects arising from driving nonlinear Josephson elements in the presence of a dissipative environment. An example of this is structural instability, or ionization  \cite{lescanne_et_al_2019,verney_et_al_2019,shillito_et_al_2022,Cohen2022Jul,Dumas2024Feb}. Transmons \cite{Koch_2007} can, under drive, transition from a state prepared within their computational manifold into a state corresponding to classical chaotic dynamics, which is now known to be a major source of infidelity in qubit readout \cite{Khezri2023Nov,Sank_2016}.

\begin{figure}[t!]
    \centering    \includegraphics[width=\linewidth]{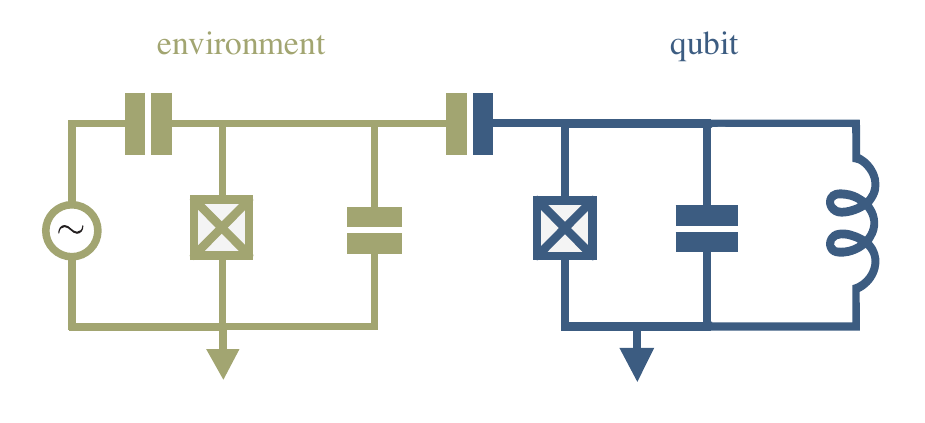}
    \caption{A lumped-element representation of a periodically driven transmon (green), capacitively coupled to another qubit (blue), shown here as a fluxonium qubit. In this work, we view the driven transmon as an environment for the circuit shown on the right, which will be modeled as a TLS.}
    \label{fig:circuit}
\end{figure}

Theoretical work so far has focused on quantifying drive-induced changes in qubit coherence \cite{boissonneault_et_al_2009,Sete2014Mar,walter_et_al_2017,Sank_2016,petrescu_et_al_2023,Thorbeck2023May}, addressing  several experiments \cite{Slichter2012Oct,vool_et_al_2014,Sank_2016,Khezri2023Nov}. Here, we shift perspective and study the noise \textit{emitted} by a transmon qubit prepared in a chaotic state. To probe this noise, we couple the chaotic transmon to another witness circuit. For our purposes we can approximate the latter with a TLS, and ask to what extent the transmon acts analogously to an infinite bath in the dynamics of the TLS [\cref{fig:circuit}]. That is, we wish to quantify the severity of transmon ionization on the degradation of coherence of its neighbors, with relevance to either multi-transmon systems \cite{Krinner_2020} or bosonic quantum error correction where transmon ancillas are used for the control and readout of bosonic qubits \cite{Lescanne2020May, putterman2024}.

In this work, we consider three models to describe the dynamics of a strongly driven transmon. The first is the full quantum model. The second corresponds to the classical limit associated with the case where the ratio between the transmon Josephson and Coulomb charging energies, $E_J/E_C$, tends to infinity. This is  the model of a driven classical pendulum, which admits both regular and chaotic regions in its phase portrait. Within the chaotic region, and for strong enough drives, we use an even simpler third model approximating the chaotic dynamics. This is a stochastic model that describes the effective diffusion of the charge variable.

Our findings are as follows. First, for a given drive and transmon frequency, and a sufficiently large $E_J/E_C$, and a large enough drive amplitude, the effect of the transmon on the short-time dynamics of the TLS is well approximated by that of a chaotic classical pendulum driving the TLS. This concerns both energy relaxation and dephasing-type effects.  Second, when considering  relaxation effects, the effective short-time dynamics can be further approximated by that induced by a simple classical stochastic noise model with a given power spectral density. We can therefore provide an analytical formula to estimate the  relaxation time ($1/T_1$) through Fermi's Golden Rule. Third, on longer time scales, quantum effects become visible through the fact that the chaotic layer of the transmon phase space corresponds to a finite-dimensional subspace of the full Hilbert space, thus limiting the analogy to a classical noise source. This leads to a saturation of the population relaxation to a plateau whose value depends on the initial state of the TLS and the number of Floquet modes in the chaotic layer. Fourth, at even stronger drives, the quantum system starts to present localization effects, and thus its dynamics deviates from the driven classical pendulum again. In particular, the induced relaxation on the TLS is weaker than predicted by a classical model.

The remainder of this paper is organized as follows. \Cref{sec:Model} presents an overview of the three models considered in this work. In \cref{sec:TLS} we consider relaxation and dephasing dynamics of a spectator TLS coupled transversally to a driven transmon. We conclude in \cref{sec:Con}. Appendices referenced in the text contain technical details.

\section{From driven transmon to diffusive stochastic process}
In this section we introduce the three models for the driven transmon used throughout this paper. We begin with the quantum mechanical description of the transmon qubit in \cref{sec:QuT}. We then define its classical limit in \cref{sec:semiclassical}, and, under certain conditions for drive strength and anharmonicity, further approximate this as a diffusive stochastic process in \cref{sec:diffusive}. 

\label{sec:Model}
\begin{figure*}[ht!]
    \centering
    \includegraphics[width=\linewidth]{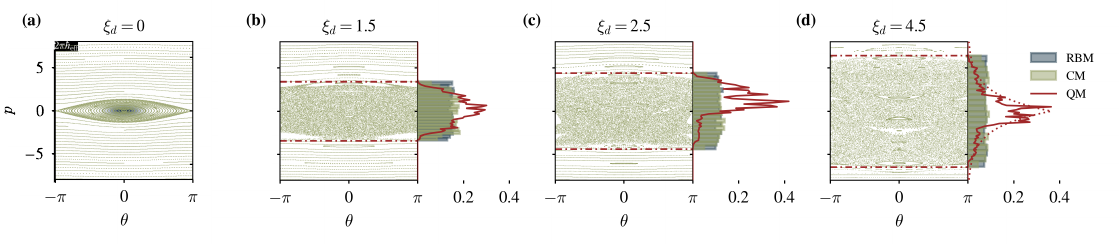}
    \caption{(a) Poincaré section of an undriven pendulum, 
    overlaid on the Husimi function of the transmon ground state at $\hbar_{\text{eff}}=0.16$, which occupies the phase space area depicted by the black rectangle. (b)-(d) Poincaré section of the driven pendulum for three different drive strengths with $\lambda=0.47$. The boundaries of the chaotic region (dash-dotted red lines) are approximately situated at $p=\pm \bar{p}$ with $\bar{p}$ approximately $\propto \xi_d$ [see \cref{App:Separatrix}]. On the right-hand side of each panel, the long-time ($t=1000T$, with $T$ period of the system) momentum distribution versus $p$ for the reflected brownian motion (RBM), and for the classical model (CM)  are uniform over the chaotic region. The classical distribution deviates from the uniform distribution close to the stability islands. The squared wavefunction of the quantum model (QM) versus $p$ averaged over $n_g$, \ie $\mathbb{E}_{n_g} |\langle p | \psi(t)\rangle|^2$, when $\lambda=0.47$ and the same value of $\hbar_{\text{eff}}$ as in (a), is close to a uniform distribution in (b), while at sufficiently large drive strengths it localizes [panel (d)]. The red dotted line in (d) gives the exponential falloff $e^{-|n|/l_n}$ of the wavefunction with the localization length $l_n$  \cref{eq:locLength}. \label{fig:p_distrib}}
\end{figure*}

\subsection{Driven transmon: quantum model}
\label{sec:QuT}
We begin by considering the quantum mechanical model of a charge-driven transmon \cite{Koch_2007} as described by the Hamiltonian  
\begin{align} \label{eq:Htr}
    \hat H_{tr}(t) = 4 E_C (\hat n-n_g)^2 - E_J \cos \hphi -   \hbar \varepsilon_d \cos(\omega_d t) \hat n
\end{align}
where $\hn$ and $\hphi$ are the Cooper pair number and the superconducting phase difference operators obeying the canonical commutation relation $[\hphi,\hn]=i$, $E_C$ and $E_J$ are the charging and Josephson energies, $\varepsilon_d$ the drive amplitude, and $\omega_d = 2\pi/T$ the drive frequency. $n_g$ is the offset charge on the superconducting island, which we will primarily use to generate statistics \cite{Cohen2023}. The driven transmon Hamiltonian \cref{eq:Htr} is one possible approximation of the circuit QED Hamiltonian in which quantum fluctuations of the readout resonator are neglected \cite{Cohen2023}. 

Under some conditions for drive strength and frequency, a transmon initially prepared in its computational manifold can be excited to higher states due to chaos-assisted transitions \cite{lescanne_et_al_2019,verney_et_al_2019,shillito_et_al_2022,Cohen2023,Dumas2024Feb}. For numerical simulations, we vary parameters about a reference set of typical circuit QED values \cite{Koch_2007,blais_et_al_2021}: $E_J/h = 29.37\ \mathrm{GHz}$ and $E_C/h = 0.2\ \mathrm{GHz}$ resulting in a plasma frequency $\omega_p/2 \pi= \sqrt{8 E_C E_J}/h= 6.85 \ \mathrm{GHz}$, anharmonicity $\alpha/h=-215\ \mathrm{MHz}$, while $E_J/E_C \approx 147$, with about $10$ cosine-potential-bound states. This ratio renders the quantum dynamics of the transmon close to that of its classical limit, to be defined below in \cref{sec:semiclassical}.

Here we are interested in the dynamics of a spectator degree of freedom, to which the transmon is further capacitively coupled as in \cref{fig:circuit}(a), and which we model as a TLS transversally coupled to the transmon, with the whole system obeying the Hamiltonian
\begin{align} \label{eq:Hfull}
    \hat{H}(t)= \hat{H}_{tr}(t)+ \frac{\hbar \omega_q}{2} \hat \sigma_z +\hbar g  \hat \sigma_x  \hat n, 
\end{align}
with $\omega_q$ the TLS transition frequency and $g$ the coupling strength, and $\hat{\sigma}_z = \ket{e}\bra{e} - \ket{g}\bra{g}$, $\hat{\sigma}_x = \ket{g}\bra{e}+\ket{e}\bra{g}$, where $\ket{g}$ and $\ket{e}$ are the ground and excited states of the TLS. Note that we do not consider couplings to environmental degrees of freedom that would otherwise produce intrinsic dissipation in addition to the unitary time evolution generated by the Hamiltonian \cref{eq:Hfull}. Instead, we aim to show that the transmon alone driven to a chaotic state could effectively act as a noise source to the spectator TLS [\cref{fig:circuit}(b)], under conditions specified below.

\subsection{Driven transmon: classical limit}
\label{sec:semiclassical}
As described above, in order to draw a comparison to a noise source, we initialize the transmon in a chaotic state. To define this initial state, we use a classical limit of the transmon Hamiltonian \cref{eq:Htr}. We write the corresponding Schrödinger equation 
\begin{align}
\begin{split}
    &i \hbar \frac{d \ket{\psi}}{dt} = \\
    &\quad \left[ 4 E_C (\hat n-n_g)^2 - E_J \cos\hphi  - \hbar \varepsilon_d \cos(\omega_d t) \hat n \right] |\psi \rangle,
    \end{split}
\end{align}
which we recast to (see \cref{App:ScLimit})
\begin{align}
\begin{split} 
    i \hbar_{\text{eff}} \frac{d |\tilde \psi\rangle}{d \tilde t}  =
    \left[   \frac{\hbar_\text{eff}^2(\hat n-n_g)^2}{2} - \lambda\cos(\hat \theta - \xi_d \sin(\tilde t)) \right] |\tilde \psi \rangle
\end{split}
\label{eq:Schr}
\end{align}
in terms of dimensionless time, plasma frequency, drive amplitude, and Planck constant
\begin{align} \label{eq:Resc}
\tilde t = \omega_d t, 
~~\lambda=\left(\frac{\omega_p}{\omega_d}\right)^2,~~
\xi_d=\frac{\varepsilon_d}{\omega_d},
~~\hbar_{\text{eff}}=\frac{8E_C}{\hbar\omega_d}.
\end{align}

Throughout this work we fix $\lambda=0.47$, which corresponds to a transmon plasma frequency below the drive frequency. \Cref{eq:Schr} has the form of a Schr\"odinger equation containing the dimensionless effective reduced Planck constant $\hbar_{\text{eff}}$. In terms of new coordinates $\{\hat \theta = \hphi,\hat p = \hbar_{\text{eff}} \hat n\}$,  the phase-space volume of a quantum state as given by Bohr-Sommerfeld quantization  \cite{PhysicsPhysiqueFizika.2.131}, $2\pi \hbar_{\text{eff}}$, varies with the microscopic parameters of the circuit. 

In the rewritten Schrödinger  \cref{eq:Schr}, we can reach the classical limit $\hbar_{\text{eff}} \to 0$, for example, by keeping the transmon parameters fixed and implementing a high-frequency drive $\hbar \omega_d \gg E_C$.  When $\hbar_{\text{eff}}=0$ we retrieve from \cref{eq:Schr} the Hamilton function of a classical driven pendulum
\begin{align} \label{eq:Hcls}
    H(\tilde t)= p^2/2-\lambda \cos\left[\theta-\xi_d \sin(\tilde t)\right]
\end{align}
where we have omitted the gate charge $n_g$ in the first term. In the classical model, this can be removed by a canonical transformation, unlike the quantum case, where it twists the periodic boundary condition of the wavefunction corresponding to the compact Josephson potential \cite{Koch_2007}. 

Given the time-periodic Hamiltonian \cref{eq:Hcls}, to easily characterize if the classical dynamics is chaotic, we will focus on Poincaré sections \cite{Zaslavskii_Sagdeev_Usikov_Chernikov_1991}  derived from Hamilton's equations [derivatives below are in terms of the rescaled time $\tilde{t}$ of \cref{eq:Resc}]
\begin{align}
    \begin{split}
      \dot \theta &=  p,  \\
      \dot p &=  -\lambda \sin\left[\theta - \xi_d \sin( \tilde t)\right].
    \end{split}
    \label{eq:HamEq}
\end{align}
In \cref{fig:p_distrib}\textbf{(a)} we show the Poincaré section of the undriven classical pendulum, containing closed orbits which corresponding to bounded oscillations of the pendulum, open orbits corresponding to full rotations, and the separatrix curve corresponding to pendulum energy $H=\lambda$. Alongside the Poincaré section we represent the Husimi function \cite{carmichael_2002} of the ground state of the undriven transmon Hamiltonian \cref{eq:Htr} in the coordinates $\{\hat \theta,\hat p \}$ at $\hbar_{\text{eff}}=0.16$. As we decrease $\hbar_{\text{eff}}$, since zero-point fluctuations $\theta_{\text{ZPF}}=\hbar_{\text{eff}}/\sqrt{\lambda}$, the standard deviation of the ground state wavefunction function shrinks in these coordinates, approaching a single classical trajectory (not shown). 

When the drive is turned on, the Poincaré section develops a chaotic region whose extent increases from the separatrix with increasing drive strength $\xi_d$, as shown in \cref{fig:p_distrib}\textbf{(b-d)}. We are interested in sufficiently strong drives that yield a nearly contiguous chaotic layer in a neighborhood of the origin without a central regular island reminiscent of the undriven dynamics. 
We can determine a rough threshold for the disappearance of the central regular island to be the drive amplitude at which the central nonlinear resonance completely overlaps with its neighbors  \cite{CHIRIKOV1979263} (see \cref{App:Separatrix}), $1-2\sqrt{\lambda |J_1(\xi_d^<)|}=0$, with $J_1$ being a Bessel function of the first kind \cite{abramowitz1964handbook}, which, for the parameters considered here, gives $\xi_d^< \approx 1.34$. We can analogously find a rough threshold for the reappearance of the central regular island to be the lowest solution to $1-2\sqrt{\lambda |J_1(\xi_d^>)|}=2\sqrt{\lambda |J_0(\xi_d^>)|}$, or $\xi_d^> \approx 3.8$. In the following \cref{sec:diffusive} we argue that in this region the phase space dynamics maps to a diffusive process.

\subsection{From classical model to reflected Brownian motion}
\label{sec:diffusive}
In this subsection, we introduce an approximation to the classical model for the driven pendulum, which we show maps to a stochastic diffusive process under certain conditions. For sufficiently high drive strength, in the regime of fast resonance crossing \cite{my009} $\xi_d \gg \max(1,\lambda)$, the stroboscopic evolution of the classical equations of motion \cref{eq:HamEq}, over finite times, is well approximated by the Chirikov standard map \cite{Chirikov:325497}
\begin{align}
    \begin{split}
     \theta_{(n+1)T} &= \theta_{nT} +  T  p_{(n+1)T}, \\
           p_{(n+1)T} &=   p_{nT} - k \sin(\theta_{nT})  \label{eq:CSM}
    \end{split}
\end{align}
with $k=2 \sqrt{\pi}  \lambda/\sqrt{\xi_d},$ and $T=2\pi$ is the period of the drive in the rescaled \cref{eq:Schr}. This map, which we explicitly rederive for completeness in \cref{app:standardMap} from the pendulum dynamics, generates unbounded chaotic dynamics for $kT>K_c$, a critical value close to 1~\cite{CHIRIKOV1979263}. In that case, the sequence $\theta_{nT}$ is apparently random and uniformly distributed in $[0,2\pi]$ and thus the momentum
\begin{align}
    p_{nT}-p_0 = -k \sum_{m =0}^{n-1} \sin(\theta_{mT})
\end{align}
apparently diffuses in $\mathbb{R}$ with a diffusion rate \cite{Chirikov:325497}
\begin{align}
    D=\frac{k^2}{2T}=\frac{\lambda^2}{\xi_d}.
    \label{eq:diffRate}
\end{align}

Thus, in order to build the third simplified model, which we consider to study the effect of the driven transmon on the coupled TLS, we consider the stochastic diffusive process 
\begin{align}
    p_t = f\left(\sqrt{D} W_t\right)
    \label{eq:stochastic}
\end{align} 
with $W_t$ a Wiener process. While the apparent diffusion of $p$ in the Chirikov standard map is unbounded, in the case of the periodically modulated pendulum chaotic orbits remain bounded~\cite{my009}  (see also \cref{fig:p_distrib}). Indeed, the chaotic orbits are roughly limited to $p\in[-\bar p,\bar p]$ with the boundary parameter for the chaotic layer $\bar p$ approximately $\propto\xi_d$ discussed in \cref{App:Separatrix}. In the simplified stochastic model~\eqref{eq:stochastic}, we consider reflecting boundary conditions whenever reaching the boundaries, leading to a reflected 1D Brownian motion (RBM hereafter) in the domain $[-\bar p,\bar p]$. The function in the right hand side of \cref{eq:stochastic} implementing this reflecting boundary condition
\begin{align}
f(r)&=\begin{cases}
r-4k\bar p & \text{if }r \in [(4k-1)\bar p,(4k+1)\bar p],\\
(4k+2)\bar{p} - r & \text{if }r \in [(4k+1)\bar p,(4k+3)\bar p],
\end{cases}
\label{eq:refBrown}
\end{align}
where $k$ runs over integers.

So far, we have introduced three models for the strongly driven transmon. The first one is the quantum mechanical model with a classical drive. We have derived from this a classical model, which approximates the quantum dynamics for small values of the effective Planck constant $\hbar_{\text{eff}}$. Finally, the stochastic model approximates the classical model for values of the drive amplitude $\xi_d \gg \max(\lambda,1).$ In the next subsection we discuss quantum suppression of diffusion, which limits the validity of these approximations for larger  drive amplitudes $\xi_d$.

\subsection{Quantum suppression of diffusion}
\label{sec:Loc}
\begin{figure}[t!]
    \centering
    \includegraphics[width=\linewidth]{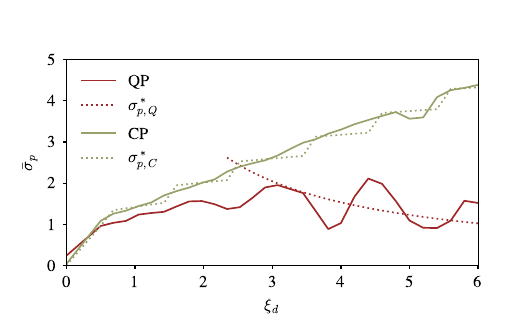}
    \caption{Asymptotic time-averaged standard deviation of momentum of the classical pendulum (CP, solid green) and of the quantum pendulum (QP, solid red, additionally averaged over offset charge $n_g$) versus the drive amplitude $\xi_d$, with $\lambda=0.47$, $\hbar_{\text{eff}}=0.16$. When the drive is off, the zero classical momentum fluctuations differ from the non-zero quantum counterpart, corresponding to zero point fluctuations.
    The linear increase of classical momentum fluctuations is well predicted by the theoretical curve $\sigma_{p,C}^*$ [\cref{eq:SDc^*}, green dotted line].  We attribute dips in the solid curves to the presence of stability islands in the mixed phase space [e.g. \cref{fig:p_distrib}\textbf{(c)}]. Localization approximately starts when $\sigma_{p,C}^*$ is larger than $\sigma_{p,Q}^*$ [\cref{eq:SDq^*}, red dotted line], i.e. $\xi_d^*\approx3.26$ from \cref{eq:xid^*}, whence QP fluctuations decrease with $\xi_d$.} 
    \label{fig:momentum fluctuations}
\end{figure}

As the drive amplitude is increased in the quantum model \cref{eq:Htr}, the motion transitions from diffusive to localized, due to quantum destructive interference effects \cite{my036,PhysRevLett.67.255}. 
This is the momentum-space analog of Anderson localization in one dimension  \cite{PhysRevA.29.1639}. An initial wavefunction, such as the ground state of the transmon, which is localized around $n=0$ (in the charge basis $\{|n\rangle\}$ formed by the eigenstates of $\hat n$), first spreads by diffusion, following the classical evolution, and then freezes into an exponentially localized probability distribution $|\psi(n)|^2 \propto e^{-|n|/l_n},$ with localization length for Cooper pair number $\hat{n}$ given by
\begin{align}
    l_n=\frac{T D}{ \hbar_{\text{eff}}^2}
    \label{eq:locLength}
\end{align}
with $T=2\pi$ the drive period introduced above, and $D$ the diffusion rate of \cref{eq:diffRate} \cite{shepelyansky1987localization, CHIRIKOV198877,PhysRevLett.56.677}. Since the diffusion rate $D$ is inversely proportional to the drive amplitude $\xi_d$, the localization length $l_n$ decreases with increasing drive amplitude. 

In \cref{fig:p_distrib}(b-d), we numerically probe dynamical localization versus drive strength. The Poincaré sections for three different drive strengths $\xi_d=1.5, 2.5$,  and $4.5$ have a chaotic layer of width $|p|<\bar{p} \sim \xi_d$ (\cref{App:Separatrix}). Along the momentum axis, we plot three distributions: the long-time momentum distribution of the diffusive process (after one thousand periods of the drive, blue), the long-time momentum distribution of the classical pendulum (green), and the long-time momentum distribution obtained from the time-evolved wavefunction of the transmon (red), the latter averaged over $n_g \in [0,0.5]$. 

To initialize the classical pendulum and the diffusive process, we sample 5000 different initial conditions from the Husimi function corresponding to the ground state of the quantum model. In \cref{fig:p_distrib}(b), for $\xi_d = 1.5$, all three distributions are nearly uniform over the chaotic layer, with deviations in the quantum and classical models that we attribute to the presence of regular islands. In panel (d), however, for the larger drive strength $\xi_d=4.5$, the squared wavefunction has an exponential decay determined by localization length $l_n$ as in \cref{eq:locLength}.

Another signature of localization appears in the behavior of long-time (asymptotic) momentum fluctuations $\bar \sigma_p$  with increasing drive amplitude $\xi_d$. In the classical case, a set of trajectories evolving from a patch in phase space included in the chaotic domain diffuse and asymptotically cover the entire chaotic domain. The associated asymptotic momentum fluctuations are determined by the standard deviation of a stochastic process uniformly distributed in $[-\bar p,\bar p]$  
\begin{align}
    \sigma_{p,C}^*=\bar p/\sqrt{3},
    \label{eq:SDc^*}
\end{align}
where $\bar p$ is the maximum momentum in the chaotic layer as estimated with Chirikov's criterion of non-overlapping resonances (\cref{App:Separatrix}).

In the quantum case, localization reduces asymptotic momentum fluctuations to \cite{PhysRevLett.67.255}
\begin{align}
    \sigma_{p,Q}^*=\hbar_{\text{eff}} l_n/\sqrt{2}.
    \label{eq:SDq^*}
\end{align}
Localization should set on when the classical fluctuations due to the border of the chaotic domain, $\sigma_{p,C}^*$, exceed the fluctuations restricted due to dynamical localization, $\sigma_{p,Q}^*$, \ie from 
\begin{align} \label{eq:xid^*}
\xi_d^*\approx(\sqrt{6}\pi \lambda^2/\hbar_{\text{eff}})^{1/2},
\end{align}
obtained upon coarsely approximating $\bar{p}\approx \xi_d$ (we do not have a closed form for the more precise estimate of $\bar{p}$, obtained numerically in \cref{App:Separatrix}).
\Cref{fig:momentum fluctuations} shows that numerically extracted asymptotic momentum fluctuations $\bar \sigma_p$ versus the drive strength $\xi_d$ for the classical (green) and quantum model (red) agree well with the theoretical estimates of \cref{eq:SDc^*} and \cref{eq:SDq^*}. Fluctuations for the quantum model reach a maximum at approximately $\xi_d^*$.

\section{Dynamics of two-level system coupled to driven transmon}
\label{sec:TLS}

\subsection{Short-time energy relaxation dynamics}
\label{subsec:T1 short}
In this section, we study the relaxation dynamics of the TLS coupled to the driven transmon. We show that the population of the TLS relaxes exponentially towards a maximally mixed state, as if the TLS were coupled to an infinite temperature bath. 

In the frame of \cref{eq:Schr}, the Hamiltonian for the TLS coupled to the driven transmon \cref{eq:Hfull} is
\begin{align}
    \begin{split}
        \frac{\hat H(\tilde t)}{\hbar_{\text{eff}}} = &\frac{\tilde \omega_q}{2} \hat \sigma_z  + \tilde g \hbar_{\text{eff}} \hat n \hat \sigma_x \\ &+\frac{\hbar_{\text{eff}}}{2}(\hat n - n_g)^2 - \frac{\lambda}{\hbar_{\text{eff}}} \cos(\hphi-\xi_d \sin(\tilde t))
    \end{split}
    \label{eq:TOTHam}
\end{align}
with dimensionless TLS transition frequency $\tilde \omega_q =  \omega_q/ \omega_d$ and coupling  $\tilde g=\hbar g/(8 E_C)$ (\cref{App:ScLimit}). In numerical simulations we use the parameters $\hbar_{\text{eff}}$  and $\lambda$ of \cref{fig:p_distrib}, and $\tilde g = 0.01$ ($g/2\pi=16\,\mathrm{MHz}$), corresponding to the circuit parameters defined under \cref{eq:Htr}. We moreover choose the frequency of the TLS to be $\tilde \omega_q = 1/\sqrt{2}$. For details on numerical simulations of the quantum model, see \cref{App:Num}.

We compare below the dynamics resulting from \cref{eq:TOTHam} to the Schrödinger equation of the TLS coupled to the classical or stochastic models, which we derive  through a semiclassical approximation
\begin{align}
        i \frac{d}{d \tilde t} | \tilde \psi \rangle = \Big[  \frac{\tilde \omega_q}{2} \hat \sigma_z + \tilde g p_{\tilde t} \hat \sigma_x \Big] |\tilde \psi \rangle,
        \label{eq:SEsemi}
\end{align}
where the evolution of the classical charge variable $p_{\tilde t}$ is generated by either the classical equations of motion for the pendulum \cref{eq:HamEq} or by the reflected Brownian motion \cref{eq:refBrown}. 

In numerical simulations, we initialize the TLS in its excited state $\ket{e}$. Furthermore, similarly to \cref{sec:Loc}, while in the full quantum case we initialize the transmon in its ground state, in the semiclassical case \cref{eq:SEsemi} we sample the initial condition from the corresponding Husimi function and average the TLS observables over the initial conditions. Moreover, since the transmon Hamiltonian in \cref{eq:TOTHam} is offset charge $n_g$-dependent, we average the trajectory of the TLS observables over $n_g$. In the limit of very small $\hbar_{\text{eff}}$ this has little impact on the result, as the large number of states in the chaotic layer leads to a single-rate exponential decay. However, for higher values of $\hbar_\textit{eff}$, TLS observables typically have oscillations that disappear upon $n_g$ averaging. Such averaging over the background charge is not necessary in the classical and stochastic models, which are $n_g$-insensitive.

\begin{figure}[t!]
    \centering \includegraphics[width=\linewidth]{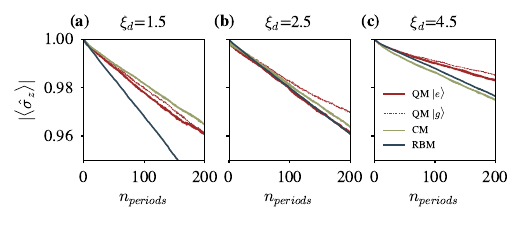}
     \caption{Short-time TLS relaxation dynamics when initialized in the excited state $\ket{e}$. We compare the full quantum model~\cref{eq:TOTHam} (solid red) with two semi-classical models~\cref{eq:SEsemi} where $p_{\tilde t}$ follows either the classical dynamics~\cref{eq:HamEq} (solid green) or the reflected Brownian motion~\cref{eq:refBrown} (solid blue). When the TLS is initialized in $\ket{g}$ (quantum model, dot-dashed line), the relaxation dynamics agrees in absolute value to the previous case. The TLS frequency $\tilde \omega_q=1/\sqrt{2}$, the coupling strength $\tilde g=0.01$,  $\lambda=0.47$, and $\hbar_{\text{eff}}=0.16$. The three plots correspond to three drive strengths exploring the various regimes of transmon chaotic dynamics (see text and \cref{fig:p_distrib}).} 
    \label{fig:T1short}
\end{figure}

We present results from our numerical simulations in \cref{fig:T1short}. We consider $\hbar_{\text{eff}}=0.16$, and three drive strengths $\xi_d=1.5, 2.5$ and $4.5$ (as in \cref{fig:p_distrib}). With these choices of parameters we can explore various chaotic regimes for the driven transmon, as discussed in \cref{sec:Model}. As we increase the drive strength from $\xi_d = 1.5$ to $\xi_d=2.5$, panels (a) and (b), we enter the fast resonance crossing regime such that the RBM better approximates the classical model, which in turn matches the quantum model. In panel (c),  the strong drive $\xi_d=4.5$ lies beyond the threshold for quantum localization, \cref{eq:xid^*}, therefore the classical and diffusive models deviate from the quantum model, which gives a slower decay for the TLS.

We now consider the case where the TLS is initialized in its ground state $\ket{g}$ instead. For a classical noise source we expect a symmetric behavior, indicative of similar excitation and relaxation rates. This apparent symmetry arises within a Fermi's Golden Rule calculation. Note that the power spectral density of the input classical noise $p_{\tilde t}$ is an even function of the frequency. In particular, for the semiclassical model where the transmon dynamics is replaced by a reflected Brownian motion, the decay or excitation rate of the TLS is given by $\tilde g^2 S_{pp}[-\tilde\omega_q]=\tilde g^2 S_{pp}[\tilde\omega_q]$, 
which is approximately (\cref{app:FGR})
\begin{align}
    \label{eq:fgr}
    \gamma_{\downarrow} =\gamma_{\uparrow}=  \frac{512 }{\pi^2} \left[\frac{\tilde g^2 \bar{p}^{4}D}{\tilde \omega_q^2\bar{p}^{4}+\pi^4D^2}+\frac{\tilde g^2 \bar{p}^{4}D}{\tilde\omega_q^2 \bar{p}^{4}/9+9\pi^4D^2}\right],
\end{align}
with $D$ the diffusion constant of \cref{eq:diffRate} and $\bar{p}$ the boundary of the chaotic layer estimated in~\cref{App:Separatrix}.

In~\cref{fig:T1short}, we show that, in the quantum model as well (for further details, see \cref{ap:DriTr}), there is a symmetry between the excitation and relaxation dynamics of the TLS (red solid and dot-dashed lines, respectively). This is usually a signature of coupling to an infinite temperature bath.

\begin{figure}
    \centering
    \includegraphics[width=\linewidth]{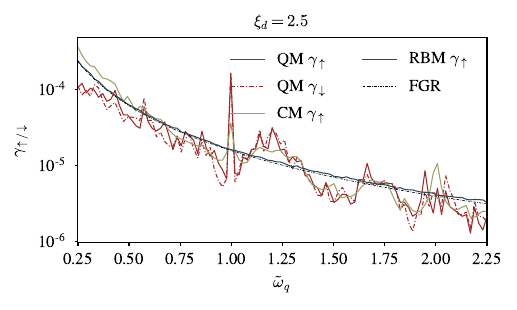}
    \caption{Decay rates $\gamma_{\uparrow}$ and $\gamma_{\downarrow}$ of the TLS, in the  setting of \cref{fig:T1short}, varying the TLS frequency $\tilde \omega_q$ in the interval $[0.25,2.25]$. We obtain the decay rate  using a linear regression model over the first $200$ periods of the $\log$ scale dynamics of $\hat \sigma_z$. The decay rates $\gamma_{\uparrow}$ and $\gamma_{\downarrow}$ are approximately the same in the quantum case, a behavior similar to that of a TLS coupled to an infinite temperature bath.}
    
    \label{fig:gamma1}
\end{figure}

The above analytical formula~\cref{eq:fgr} captures the order of magnitude and general trend of the relaxation rate  $\gamma_{\downarrow/\uparrow}$, shown in \cref{fig:gamma1} in log scale against the TLS frequency $\tilde \omega_q$ at $\xi_d=1.5$. More precisely,  the full quantum model and the semiclassical model with the classical chaotic pendulum driving the TLS have similar relaxation rates over this wide range of TLS frequencies. Nonetheless, the semiclassical model with the driven transmon replaced by the reflected Brownian motion, for which the relaxation rate is very well estimated by the analytical formula~\cref{eq:fgr}, deviates from the other two around resonances between the drive and the TLS, marked with crosses on the plot. At these resonances, the transmon drive (or its harmonics) directly drives the TLS, resulting in a dressing of its spectrum that obscures relaxation processes.

In summary, we have shown that, on short time scales, the strongly driven transmon induces an exponential  relaxation of the TLS, thus behaving as an infinite temperature bath. We have also shown that, in a driving regime correspoding to fast crossing and before quantum localization, the strongly driven transmon can be replaced by a classical stochastic noise source, which we have characterized as a reflected Bronwnian motion in a bounded domain. One limit to the analysis in this subsection is the fact that in the quantum model the chaotic subspace of the transmon spectrum is finite-dimensional, and there is therefore a recurrence time after which part of the energy passed from TLS to the driven transmon is coherently returned to the TLS and vice versa. We analyze this in detail in the next section.

\subsection{Long-time relaxation dynamics}
\label{subsec:T1 long}
In this section, we analyze the dynamics of the TLS on longer time scales and show that, contrary to the semiclassical models where the TLS converges to the maximally mixed state, in the quantum model, the populations of the TLS state tend to values that are dependent on both their initial-state populations and the number of quasi-energy levels in the chaotic layer. 

\begin{figure}[t!]
    \centering
    \includegraphics[width=\linewidth]{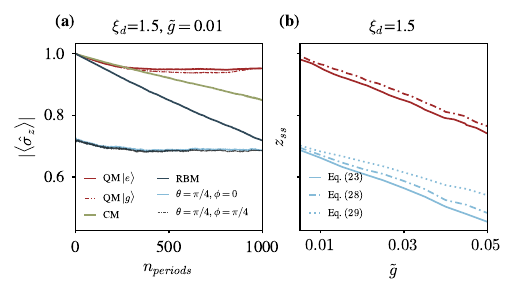}
    \includegraphics[width=0.55\linewidth]{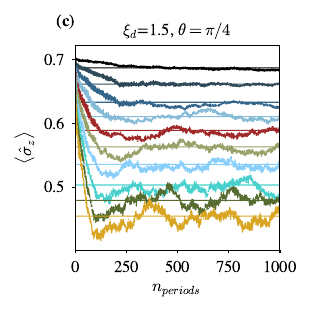}
    \caption{(a) Same conventions as in of \cref{fig:T1short}(b) for a longer time window, showing the plateau of $|\langle \sigma_z \rangle|$ decay coupled to the transmon (QM). For initialization in the excited state, $\theta = 0$, QM is shown in red, CM in green, and RBM in blue. The latter two do not exhibit a plateau. The QM plateau value in absolute value is similar if the qubit is initialized in the ground state $\theta=0$ (red dot-dashed). For comparison, QM results for $\theta=\pi/4$ are shown in light and dot-dashed blue for two different values of the phase $\phi$. (b) Value of the plateau as a function of the coupling strength $\tilde g$ from \cref{eq:dressed2} averaged over one period (solid), along with its approximation by \cref{eq:approxPlateau}, with IPR cut at $0.3$ (dashed), when the TLS is initialized in the excited state $\theta=0$ (red) or with initial angles $\theta=\pi/4\, \phi=0$  (light blue). Estimates from \cref{eq:dressed3} and \cref{eq:dressed4} are shown in dot-dashed and dotted curves.  (c) Long-time dynamics of $\sigma_z$, for ten uniformly spaced values of $\tilde g$, between  $\tilde g=0.005$ (black) and  $\tilde g=0.05$ (yellow), and $\theta=\pi/4$. We use the nearly horizontal curves obtained from \cref{eq:dressed2} to estimate the full plateau value, obtained from \cref{eq:dressed}. }
    \label{fig:T1long}
\end{figure}

In \cref{fig:T1long}\textbf{(a)}, we plot the expectation value $\langle \hat \sigma_z \rangle$ versus time for $1000$ periods of the drive, and averaged over 50 values of $n_g$ in the interval $[0,0.5]$, at drive strength $\xi_d=1.5$.  We have checked numerically (not shown), that both semiclassical models converge to the maximally mixed steady state represented by the TLS reduced density matrix $\rho_{ss}=\ket{g}\bra{g}+\ket{e}\bra{e}$, whereas the steady state for the quantum model,  $\rho_{ss}=p_g^{\infty}\ket{g}\bra{g}+p_e^{\infty}\ket{e}\bra{e}$, depends on the populations for the initial TLS state $\ket{\psi_0}=c_g^0\ket{g}+c_e^0\ket{e}$, with dependence on the initial phase $\mathrm{arg}(c_e^0/c_g^0)$ vanishing in the classical limit of small $\hbar_{\textit{eff}}$. We observe that there is a symmetry between the dynamics with the TLS initialized in state $\ket{g}$ or $\ket{e}$. In both cases, the absolute value of the expectation value of $\sigma_z$, evaluated at each driving period, converges to approximately the same value $z_{ss}$. This value, for which we provide an estimate below in \cref{eq:plateauPred}, is entirely determined by the Floquet spectrum of the coupled TLS-transmon Hamiltonian.

More precisely, we observe that initializing the TLS-transmon system in the 
superposition state 
$\ket{\psi_0}=\left(\sin\frac{\theta}{2}\ket{g}+e^{i\phi}\cos\frac{\theta}{2}\ket{e}\right) \otimes \ket{0}$,  
the steady state is given by 
\begin{align}
\label{eq:ssT1}
\rho_{ss}\approx\frac{1}{2}\left[1-\cos(\theta) z_{ss}\right]\ket{g}\bra{g}+\frac{1}{2}\left[1+\cos(\theta)z_{ss} \right]\ket{e}\bra{e},
\end{align}
which is determined by the populations of the TLS states $\ket{g}$ and $\ket{e}$ in the initial state, but not by their relative phase $\phi$.

To put the observation above on more solid grounds, we start by expanding  the solution to the time-dependent Schr\"odinger equation
\cref{eq:TOTHam} over Floquet modes \cite{Grifoni1998DrivenQT} of the driven Hamiltonian
\begin{align}
 \ket{\psi(t)}= \sum_{\alpha}c_{\alpha} e^{-i\eps_{\alpha}t}|\phi_{\alpha}(t)\rangle,
 \label{eq:floquetSol}
\end{align}
where $|\phi_{\alpha}(t)\rangle$ are the $T$-periodic Floquet modes in the first Brillouin zone, $\eps_\alpha$ are the corresponding quasienergies, and $c_\alpha = \langle \phi_{\alpha}(0)|\psi_0\rangle$ are overlaps with the initial  state. Then 
\begin{align}
   \langle \psi(t)| \hat \sigma_z |\psi(t) \rangle &= \sum_{\alpha, \beta} c_{\alpha}^*c_{\beta} \ e^{-i(\varepsilon_{\beta}-\varepsilon_{\alpha})t} \langle \phi_{\alpha}(t)| \hat \sigma_z |\phi_{\beta}(t)\rangle. 
\label{eq:dressed}
\end{align} 
Over long times the cross product terms in the above sum average out and we have
\begin{align}
   \langle \psi(t)| \hat \sigma_z |\psi(t) \rangle \approx \sum_{ \alpha} |c_{\alpha}|^2 \langle \phi_{\alpha}(t)| \hat \sigma_z |\phi_{\alpha}(t)\rangle,
\label{eq:dressed2}
\end{align}
which we have verified agrees well with the time dependence within the plateau obtained from \cref{eq:dressed} upon averaging over gate charge $n_g$ (see \cref{fig:T1long}\textbf{(c)}). 

Noting that the right-hand side of~\cref{eq:dressed} is $T$-periodic,  we write each Floquet mode as
\begin{align}
    \ket{\phi_\alpha(n T)}=r_{\alpha,g}\ket{g}\otimes\ket{\tilde \phi_{\alpha,g}}+r_{\alpha,e}\ket{e}\otimes\ket{\tilde \phi_{\alpha,e}}, \label{eq:phialpha}
\end{align}
where $\{|\tilde\phi_{\alpha,g}\rangle\}$ and $\{|\tilde\phi_{\alpha,e}\rangle\}$  are two sets of kets  belonging to the Hilbert space of the transmon. Note that these two sets become the set of Floquet modes of the transmon, and therefore a basis for that Hilbert subspace, only when the TLS-transmon coupling vanishes $g=0$. As the coupling of the TLS to the transmon is through the $\hat{\sigma}_x$ Pauli operator, one can choose the phases of the transmon states $|\tilde \phi_{\alpha,g}\rangle$ and $|\tilde\phi_{\alpha,e}\rangle$ such that, for all $\alpha$, the quantities $\langle\tilde\phi_{\alpha,g}|\tilde\phi_{\alpha,e}\rangle$, $r_{\alpha,g}$, and $r_{\alpha,e}$ are real. By the orthonormality of the Floquet modes $\{ \ket{\phi_\alpha(k T ) }\}$ we have $r_{\alpha,g}^2 + r_{\alpha,e}^2  = 1$. Using now \cref{eq:phialpha} into~\cref{eq:dressed2},
\begin{align}
\label{eq:dressed3}
\langle \psi(nT)| \hat \sigma_z |\psi(nT)\rangle\approx\sum_\alpha z_\alpha l_
\alpha,
\end{align}
where $z_\alpha = r_{\alpha,e}^2 - r_{\alpha,g}^2$, and 
\begin{align}
\begin{split}
\label{eq:dresscoeffs}
l_\alpha =& \sin(\theta)
r_{\alpha,g}r_{\alpha,e}\Re(e^{-i \phi} d_{\alpha,g}d_{\alpha,e}^*)\\&+
\sin^2 \left(\frac{\theta}{2}\right)\, r_{\alpha,g}^2|d_{\alpha,g}|^2
+\cos^2\left(\frac{\theta}{2}\right)\, r_{\alpha,e}^2|d_{\alpha,e}|^2,
\end{split}
\end{align}
with $d_{\alpha,g}=\langle\tilde\phi_{\alpha,g}|0\rangle$ and $d_{\alpha,e}=\langle\tilde\phi_{\alpha,e}|0\rangle$. Since the transmon ground state $\ket{0}$ is within the chaotic region, its overlaps with regular states are zero. Therefore, in the sum~\cref{eq:dressed3}, we can keep only the chaotic states, and hereafter the sums run only over those $\alpha$ that are associated to chaotic Floquet modes. In practice, when comparing numerical results in \cref{fig:T1long}, we select as chaotic those states which have an inverse participation ratio \cite{Kramer1993Dec} (IPR) lower than $0.3$ over the decoupled $g=0$ eigenbasis  of the Hamiltonian of \cref{eq:TOTHam}.

Restricting thus to states $|\tilde{\phi}_{\alpha,g}\rangle$ and $|\tilde{\phi}_{\alpha,e}\rangle$ within the chaotic layer of the transmon,
in the limit of small $\hbar_{\text{eff}}$, their overlaps on the transmon ground state $\ket{0}$ should be similar and the overlap phases random~\cite{my153}. We can estimate 
\begin{align}
    \begin{split}
       d_{\alpha,g} &\approx \frac{e^{i\theta_{\alpha,g}}}{\sqrt{N_{ch}}},\;  d_{\alpha,e} \approx \frac{e^{i\theta_{\alpha,e}}}{\sqrt{N_{ch}}}, \\
       \Re(e^{-i\phi}d_{\alpha,g}d_{\alpha,e}^*) &\approx  \frac{\cos(\phi) \cos(\theta_{\alpha})}{N_{ch}}, \label{eq:dapprox}
    \end{split}
\end{align}
where $N_{ch}$ stands for the number of transmon chaotic Floquet modes, and we have used the fact that $\theta_{\alpha,g}$,  $\theta_{\alpha,e}$, and $\theta_{\alpha} \equiv \theta_{\alpha,g}-\theta_{\alpha,e}$ only take values $0$ or $\pi \,\mathrm{mod}\, 2\pi$ since the overlaps $d_{\alpha,g/e}$ are real. Then~\cref{eq:dressed3} leads to
\begin{align}
\begin{split}
&\langle \psi(nT)| \hat \sigma_z |\psi(nT)\rangle \approx \\ 
&\quad + \frac{\sin^2\frac{\theta}{2}}{N_{ch}}\sum_\alpha r_{\alpha,g}^2 z_{\alpha}  
+\frac{\cos^2\frac{\theta}{2}}{N_{ch}}\sum_\alpha r_{\alpha,e}^2 z_{\alpha}  \\
&\quad +\frac{\sin\theta \cos\phi}{N_{ch}} \sum_\alpha \cos(\theta_\alpha) r_{\alpha,g}r_{\alpha,e}z_{\alpha}.
\label{eq:approxPlateau}
\end{split}
\end{align}

In the limit of many chaotic levels, $\hbar_{\textit{eff}} \to 0$, the $\cos\theta_\alpha$ factors yield a sign $\pm$ that is random and uniformly distributed as $\alpha$ is varied, and 
the third term in the above sum averages to 0. Any residual dependence of the plateau value on the phase of the initial wavefunction $\phi$, as given by the last term above, is thus expected to vanish in the classical limit. Furthermore, noting that $1 \pm z_\alpha = 2 r_{\alpha,e/g}^2$, 
\begin{align}\label{eq:dressed4}
\langle \psi(nT)| \hat \sigma_z |\psi(nT)\rangle&\approx \frac{1}{2N_{ch}}\sum_\alpha z_\alpha+\frac{\cos(\theta)}{2N_{ch}}\sum_\alpha |z_\alpha|^2\notag\\
&=\frac{\cos(\theta)}{2N_{ch}}\sum_\alpha |z_\alpha|^2,
\end{align}
where, for the last equality, we have used the fact that
\begin{align}
    \sum_\alpha r_{\alpha,g}^2=\sum_\alpha r_{\alpha,e}^2=N_{ch}.
\end{align}
A similar calculation shows that the other Pauli operators $\hat\sigma_{x,y}$ average to 0. Thus, we find the steady state of ~\cref{eq:ssT1} with 
\begin{align}
z_{ss}\approx\frac{1}{{2}N_{ch}}\sum_\alpha |z_\alpha|^2={\half}\text{Var}_\alpha(z_\alpha).
\label{eq:plateauPred}
\end{align}

To now compare all of these approximations, in \cref{fig:T1long}(b) we show that the plateau value decreases with increasing TLS-transmon coupling $\tilde{g}$ taken in the interval $(0,0.05]$. This can be explained by the fact that, as the coupling $\tilde{g}$ increases, the transmon Floquet modes become more hybridized with the qubit state such that, for more values of $\alpha$, $z_\alpha$ approaches 0. In the same plot, we numerically observe that \cref{eq:plateauPred}, via \cref{eq:dressed4}, provides a good approximation of the plateau value. As argued above, we expect the difference between \cref{eq:approxPlateau} and \cref{eq:dressed4} to vanish as the classical limit is approached. An analysis of the limit of the variance in \cref{eq:plateauPred} when the number of chaotic levels increases could be performed within random matrix theory \cite{Haake}, but falls out of the scope of the present paper. We expect to retrieve in the limit of $N_{ch}\rightarrow\infty$  the behavior of a TLS coupled to an infinite bath at infinite temperature.  

\subsection{Dephasing dynamics}
\label{sec:T2effects}
Pure dephasing under a weak transverse coupling of the TLS to a noise source can only be described at fourth order in perturbation theory in the coupling, leading to an effective longitudinal coupling to the square of the noise~\cite{Reichman-Silbey-96,Makhlin-Shnirman-2003}. Pure dephasing of the TLS is then dominated by the low-frequency components of this noise. This indicates that approximations of the noise source that neglect low-frequency components should not provide a correct estimation of the dephasing dynamics. 

\begin{figure}
    \centering
    \includegraphics[width=\linewidth]{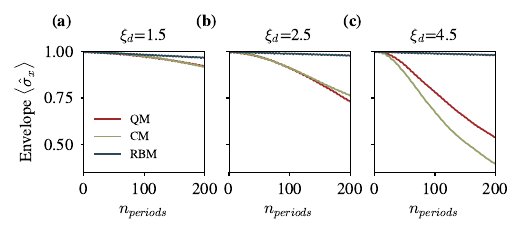}
    \caption{(a)-(c) Upper envelope of the damped oscillations experienced by the average value $\langle \sigma_x \rangle$, oscillating at the Lamb shifted frequency of the TLS. The parameters are the same as in  \cref{fig:T1short}, with initial conditions as described in \cref{sec:T2effects}. 
    }
    \label{fig:T2}
\end{figure}

We confirm this through numerical simulations in~\cref{fig:T2}. With initial condition  $\theta=\pi/2,\phi=0$, we plot the time evolution of $\langle \sigma_x \rangle$ under the Schr\"odinger \cref{eq:TOTHam} and \cref{eq:SEsemi}. The average value $\langle \sigma_x \rangle$ undergoes damped oscillations at the Lamb shifted frequency of the TLS. We demodulate these oscillations by plotting their upper envelope, so as to compare the phase relaxation dynamics in our three different models. While the coupling to the classical driven pendulum seems to agree with the dephasing dynamics of the full quantum model, the semi-classical model with the RBM as the noise source indicates a significantly slower dephasing, at a rate close to $(\gamma_\uparrow+\gamma_\downarrow)/2\approx \gamma_\downarrow$. We attribute this discrepancy to the fast-crossing approximation of the classical driven pendulum by the standard map (detailed in~\cref{app:standardMap}), leading to the approximation in the chaotic regime by a RBM, is only valid over a finite number of periods. Hence, in the Fourier domain, the two noise models only behave similarly close to the TLS frequency, as we have shown in the previous sections. We conclude that the RBM should not capture well the low-frequency behavior of the classical or quantum driven pendulum. 

\section{Conclusion}
\label{sec:Con}
We have studied a transmon driven into chaotic motion and have shown that, in a large range of parameters, the signal at the charge port of this transmon can be interpreted as a classical noise source. To argue this, we have coupled the transmon to a spectator TLS which acts as a noise spectrometer. We have found that under certain conditions for the microscopic parameters describing the transmon and the drive, the short-time dynamics of the TLS in the presence of the transmon is well approximated by that of a TLS coupled to a chaotic classical pendulum, both in terms of energy relaxation and dephasing.  Furthermore, when considering relaxation effects only, the short-time dynamics can be further approximated by that induced by a reflected brownian motion with a given power spectral density, which yields analytical formulae for the relaxation and excitation rates of the TLS. 

On longer time scales, quantum effects kick in. This occurs because the chaotic layer of the transmon phase space corresponds to a finite-dimensional subspace of its Hilbert space, resulting in a sparse spectral density giving rise to revival effects in the dynamics, and thus limiting the analogy to a classical noise source. In particular, we numerically and analytically observed that this leads to a saturation of the population relaxation to a plateau whose value depends on the initial state of the TLS and the number of Floquet modes in the chaotic layer. Finally, for stronger drives, the TLS witnesses the dynamical localization of the transmon, exhibiting a population relaxation that is weaker than predicted by the semiclassical model. 

Noting that in this work we have abstracted the spectator degree of freedom to a TLS, our study is potentially relevant to understanding a class of chaos-related cross-talk in multi-transmon systems, as well as to predicting chaos-induced suppression of coherence in some circuit-QED implementations of bosonic quantum error correcting codes, where the transmon is used for Wigner tomography of the memory register \cite{lescanne_et_al_2019,reglade_quantum_2024}. Understanding the latter is the subject of future work.

\section*{Acknowledgments}
We thank Joachim Cohen for a prior collaboration that informed this work, and Dima Shepelyansky for helpful discussions. This project was funded by ANR grant OCTAVES (ANR-21-CE47-0007). The authors also acknowledge funding from the Plan France 2030 through the project ANR-22-PETQ-0006. Simulations were performed on the CLEPS high-performance computing cluster hosted by Inria Paris.

\begin{appendix}
\section{Change of coordinates}
\label{App:ScLimit}
In this appendix we detail the change of coordinates that leads to the dimensionless Hamiltonian of \cref{eq:Schr}. The Schrödinger equation describing the time dependence of the state of the driven transmon reads 
\begin{align}
    \begin{split}
    i \hbar \frac{d}{dt} | \psi \rangle = \Big[ 4 E_C (\hat n-n_g)^2 - E_J \cos(\hphi) \\- \hbar \varepsilon_d \cos(\omega_d t) \hat n \Big] |\psi \rangle.
    \end{split}
\end{align}
The displaced states $|\tilde{\psi}(t)\rangle = e^{-i \xi_d \sin(\omega_d t) \hat n}  |\psi(t) \rangle $, with $\xi_d=\eps_d/\omega_d$, obey the  Schr\"odinger equation with rescaled time $\tilde t = \omega_d t$
\begin{align}
\begin{split}
    i \hbar \omega_d \frac{d}{d\tilde t} | \tilde \psi \rangle = \Big[ 4 E_C (\hat n-n_g)^2 - E_J \cos(\hphi -\xi_d \sin(\tilde t))  \Big] |\tilde \psi \rangle.
    \end{split}
    \label{eq:Schr1}
\end{align} 
We renormalize the Hamiltonian by multiplying both sides of \cref{eq:Schr1} by $8E_C/(\hbar \omega_d)^2$ and we obtain
\begin{align}
\begin{split}\label{ApEq:SET}
    i \hbar_{\text{eff}} \frac{d}{d\tilde t} | \tilde \psi \rangle = \Big\{ \frac{[\hbar_{\text{eff}}(\hat n-n_g)]^2}{2}  - \lambda \cos(\hphi -\xi_d \sin(\tilde t))  \Big\} |\tilde \psi \rangle
    \end{split}
\end{align}
with $\lambda =\frac{8E_JE_C}{(\hbar \omega_d)^2}$ and $\hbar_{\text{eff}}=\frac{8E_C}{\hbar \omega_d}$, which plays the role of the effective Planck constant. Written in this form, we can deduce the Hamiltonian in the classical limit ($\hbar_{\text{eff}}\rightarrow 0$), by replacing the  dimensionless momentum $\hat p = \hbar_{\text{eff}} \hat n$
and position $\hphi$ operators with their classical counterparts.

Finally, in this displaced frame, the Hamiltonian of the full system, consisting of the transverse coupling of a TLS to the transmon, is, in the same units as the transmon Hamiltonian on the right-hand side of \cref{ApEq:SET},
\begin{align}
    \begin{split}
        \frac{\hat H(\tilde t)}{\hbar_{\text{eff}}} = &\frac{\tilde \omega_q}{2} \hat \sigma_z  + \tilde g \hbar_{\text{eff}} \hat n \hat \sigma_x \\ &+\frac{\hbar_{\text{eff}}}{2}(\hat n - n_g)^2 - \frac{\lambda}{\hbar_{\text{eff}}} \cos\left[\hphi-\xi_d \sin(\tilde t)\right],
    \end{split}
    \label{eq:HamTot}
\end{align}
where $\tilde \omega_q =  \omega_q/ \omega_d$ and $\tilde g=\frac{\hbar g}{8 E_C}$, and $g$ the light-matter coupling, which is \cref{eq:TOTHam} of the main text. Here, as well, the semiclassical model is obtained by replacing the dimensionless transmon  quadratures $\hat p = \hbar_{\text{eff}} \hat n$ and $\hat\varphi$ by their classical counterparts, while keeping the quantum description of the TLS.

\section{Analysis of semiclassical model}
\label{app:standardMap}

\subsection{From driven pendulum to diffusive process}
In this section, we provide an alternative argument to Chirikov's  \cite{Chirikov:325497} to approximate the equations of motion of the classical driven pendulum (\ref{eq:HamEq}) by a diffusive process. We start with a change of variables to
\begin{align}
    \psi \equiv \frac{1}{\xi_d}\theta - \sin( \tilde{t} ). \label{eq:Psi_new}
\end{align}
We are interested in the limit $\xi_d \gg 1$. To emphasize this, we introduce a small parameter $\eps=\frac{1}{\xi_d} \ll 1$. We also assume $\lambda$ is of order $1$. In terms of the new variables $\psi$ and $p$, \cref{eq:HamEq} become
\begin{align}
    \begin{split}
      \dot \psi &=  \epsilon p  - \cos \tilde{t},  \\
      \dot p &=  -\lambda \sin \frac{\psi}{\epsilon},
    \end{split}
    \label{eq:EOMclspsi}
\end{align}
For large $p \gg \xi_d=1/\eps \gg 1$, $\dot \psi$ is large, and thus $\dot p$ can be averaged to $0$ by the second equation above, which leaves $p$ almost constant. The same reasoning works for $p \ll -\xi_d$. This corresponds to the regular regions of phase space.

\begin{figure}[t!]
    \centering
    \includegraphics[width=\linewidth]{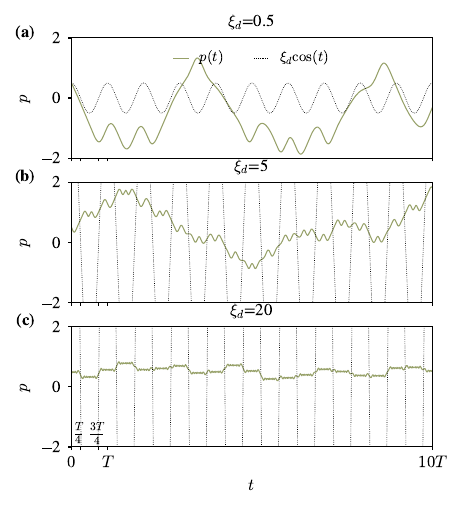}
    \caption{\textbf{(a)}-\textbf{(c)} Momentum dynamics which shows a crossing of the resonance when it intersects the resonant frequency $\omega_r=\xi_d \cos(\tilde{t}_r)$. In the fast crossing regime $\xi_d \gg 1,\lambda$ (\textbf{(b)}-\textbf{(c)}), the resonance is crossed twice per period, at $t \approx n T + T/4$ or $t \approx nT + 3T/4$. }
    \label{fig:resCrossing}
\end{figure}

We are interested in the region $\epsilon |p| \leq 1$. In this case $\dot \psi$ regularly crosses zero according to the first \cref{eq:EOMclspsi}, and these events are called resonances. When far from a resonance, $\dot p$ is again a highly oscillating function with zero average. However, close to a resonance, $\dot p$ stops to oscillate and a non-trivial time evolution of $p$ can happen. We confirmed our remarks here by numerically integrating the equations of motion (\ref{eq:HamEq}) in \cref{fig:resCrossing}.

To analytically calculate the time evolution, we make a stricter assumption than in the previous paragraph. Hereafter suppose that $\epsilon |p| \ll 1$. Note that in this case the resonances occur at times $\tilde{t}_n \approx n \pi - \frac{\pi}{2}$. Note also that at a resonance $\ddot \psi$ is approximately $(-1)^{n+1}$, and this regime is called fast-crossing as $\eps \dot p$ is negligible compared to $\sin \tilde t_n$ .

Now we approximate the jump of $p$ during the resonance crossing at $\tilde{t}=\tilde{t}_1$. We denote  $\psi_n=\psi(\tilde{t}_n)$ for $n \geq 1$, and we can write
\begin{align}
    \psi(\tilde{t})-\psi_1=-\sin(\tilde{t})+\sin(\tilde{t}_1) + O((\tilde{t}-\tilde{t}_1)\eps), 
\end{align}
for $\tilde{t} \in (\tilde{t}_0,\tilde{t}_2)$. Taking $0 <\delta \tilde{t}< \pi$, we consider
\begin{align}
    \begin{split}
    &p(\tilde{t}_1+\delta \tilde{t})-p(\tilde{t}_1-\delta \tilde{t})\\
    &=-\lambda \int_{\tilde{t}_1-\delta \tilde{t}}^{\tilde{t}_1+\delta \tilde{t}}  \sin \frac{\sin(\tilde{t}_1)-\sin(\tilde{s})+\psi_1+O((\tilde{s}-\tilde{t}_1)\eps)}{\eps} d\tilde{s}\\
    &=-\lambda\int_{-\delta \tilde{t}}^{\delta \tilde{t}}  \sin \frac{\sin(\tilde{t}_1)-\sin(\tilde{u}+\tilde
    {t}_1)+\psi_1+O(\tilde{u}\eps)}{\eps} d\tilde{u}\\
    &=-\lambda \int_{-\delta t}^{\delta t} \Im\left[ e^{i\frac{f(\tilde u)}{\eps}} e^{i\left[\frac{\psi_1}{\eps}+O(\tilde u)\right]}\right]d\tilde{u},
    \end{split}
\end{align}
with $f(\tilde{u})=-\sin(\tilde{u}+\tilde{t}_1)+\sin(\tilde{t}_1)$.

We now introduce $\tilde{u}_1=\tilde{t}_1-\pi/2$, the unique solution of $f'(\tilde{u})=0$ on the integration domain. A direct application of the stationary phase approximation gives for  $0<|\tilde{u}_1| <\delta \tilde{t} <\pi$, and $G(\tilde{u})=e^{i\left[\frac{\psi_1}{\eps}+O(\tilde{u})\right]}$
\begin{align}
\label{eq:SPA}
    \begin{split}    
    &\int_{-\delta \tilde{t}}^{\delta \tilde{t}} e^{i\frac{f(\tilde{u})}{\eps}} G(\tilde{u}) d\tilde{u} \\
    &=
    G\left(\tilde{u}_1\right) e^{i\frac{ f\left(\tilde{u}_1\right)}{\eps}+i \frac{\pi}{4} \operatorname{sign} f^{\prime \prime}\left(\tilde{u}_1\right) }\sqrt{\frac{2 \pi\eps}{\left|f^{\prime \prime}\left(\tilde{u}_1\right)\right|}}+o\left(\eps^{\half}\right).
    \end{split}
\end{align}
Using $\tilde{u}_1\approx 0$, we get
\begin{align}
   p(\tilde{t}_1+\delta \tilde{t})-p(\tilde{t}_1-\delta \tilde{t}) \approx -\lambda \sqrt{2\pi \eps}\sin\left(\frac{\psi_1}{\eps} + \frac{\pi}{4}\right).
\end{align}
At arbitrary $n$, we analogously find
\begin{align}
    \begin{split}
        \Delta p (\tilde{t}_n) \equiv& p(\tilde{t}_n+\delta \tilde{t})-p(\tilde{t}_n-\delta \tilde{t}) \\ \approx& -\lambda \sqrt{2\pi \eps}\sin\left(\frac{\psi_n}{\eps} + (-1)^{n+1} \frac{\pi}{4}\right),
    \end{split}
\end{align}
with the only difference being the sign of the phase shift in the argument of the sine, which depends on the direction in which the resonance is crossed, as shown in \cref{eq:SPA}. This sign can be determined using $\ddot{\psi}(\tilde t_n) \approx (-1)^{n+1}$.

As one period of the drive contains two consecutive resonances, the momentum change is, for $n \geq 1$,
\begin{align}
    \begin{split}
    &\Delta p (T) \equiv 
    \Delta p(\tilde{t}_{2n-1})+\Delta p(\tilde{t}_{2n}) 
    \\& = -\lambda \sqrt{2\pi \eps} \left[\sin\left(\frac{\psi_{2n-1}}{\eps} + \frac{\pi}{4} \right) +\sin\left(\frac{\psi_{2n}}{\eps} - \frac{\pi}{4} \right) \right]. 
    \end{split}
\end{align}

Following \cite{my009}, we note that the phases at the crossing of a resonance, \(\vartheta_n \equiv \psi_n/\varepsilon\) modulo \(2\pi\) can be reasonably approximated by independent and identically distributed (IID) random variables with a uniform distribution over \([0, 2\pi)\). We recall that \(\mathbb E(\sin \vartheta_n)= 0\) and the variance \(\mathbb E(\sin^2 \vartheta_n ) = 1/2\), if $\vartheta_n$ follows a uniform distribution over \(\theta \in [0, 2\pi)\).

Given that the phase shifts \(\pm \pi/4\) do not alter the distribution of \(\vartheta_n\), we find that \(\Delta p(T)\) has a zero mean. The variance of \(\Delta p(T)\) is thus given by
\begin{align}
\label{eq:diffusion_rate}
    \mathbb{E}( \Delta p(T)^2 ) = \frac{2\pi \lambda^2}{\xi_d}.
\end{align}

While we have obtained a stochastic process with IID random jumps, we can perform an explicit calculation to approximate this process with a continuous diffusion process. The diffusion rate is determined by \cref{eq:diffusion_rate}
\begin{align}
    D_p = \frac{\mathbb{E}(\Delta p(T)^2)}{T} =  \frac{\lambda^2}{\xi_d}.
\end{align}

In concluding this section, note that the analytical derivation presented in this chapter can only be valid on timescales short enough such that our assumption $\eps p \ll 1$ remains valid, that is, on timescales over which the variance of the momentum does not exceed the chaotic layer,  $\tilde{t} \lesssim \xi_d^2/D_p = \xi_d^3/\lambda^2$. This defines a low-frequency, long timescale cutoff beyond which the RBM should not be expected to well approximate the classical dynamics. This is consistent with the fact that the RBM derived under the assumptions presented in this section can capture TLS relaxation, but not dephasing, as shown in the main text.

\subsection{Estimating the size of the chaotic layer}
\label{App:Separatrix}

In this section, we recall and apply the Chirikov criterion of overlapping resonances. This heuristic criterion proposed in \cite{chirikovResonanceProcessesMagnetic1960} (see also \cite{CHIRIKOV1979263}), states that for a deterministic, time-dependent Hamiltonian system, the chaotic regions align with areas where resonances overlap. More precisely, we start with the Jacobi–Anger expansion of the time-dependent Josephson potential term in \cref{eq:Hcls}
\begin{align}
\notag
    H(\tilde t)&=p^2/2-\lambda\cos(\theta - \xi_d \sin(\tilde t)) \\
    &=p^2/2-\lambda \sum_{m=-\infty}^{\infty} (-1)^m J_m(\xi_d) \cos(\theta+m \tilde t)
    \label{eq:JA}
\end{align}
with $J_m(\xi_d)$ the $m^{\textit{th}}$ Bessel function of the first kind \cite{Abramowitz_Handbook_1964}. The \textit{principal} resonance refers to the resonance of the reduced Hamiltonian $H_0=p^2/2- \lambda J_0(\xi_d) \cos(\theta)$, \ie the separatrix which is given by $\{(p,\theta)\mid H(p,\theta)=\lambda J_0(\xi_d)\}$.

More generally, to obtain the $m^{th}$ resonance, we first consider the change of variable $\psi_m(\tilde{t}) \equiv \theta + m \tilde{t}$ and we focus on the time-independent part (in this moving frame) of the Hamiltonian, namely 
\begin{align}
    \begin{split}
        H_m &=    (p_m+m)^2/2 - (-1)^m \lambda J_m(\xi_d) \cos(\psi_m). 
    \end{split}
\end{align}

On the phase plane we can thus identify the separatrix corresponding the $m^{th}$ resonance. It is given by $\{ (p,\psi_m) \mid H_m(p,\psi_m) = \lambda |J_m(\xi_d)|\}$, which leads to the phase space curves 
\begin{align}
\label{eq:mthsep}
    p_m^{\pm}(\psi_m) =-m \pm \sqrt{2 \lambda |J_m(\xi_d)|(1+\cos\psi_m )}  
\end{align}
of maximum width $\Delta p_m = 4 \sqrt{\lambda |J_m(\xi_d)|}$ attained at $\psi_m =0$. Although the number of resonances in the set is infinite, one for each integer $m$, the widths of the separatrices $\Delta p_m $ drop rapidly for $|m| \gtrsim \xi_d$. 

\begin{figure}[t!]
    \centering
    \includegraphics[width=\linewidth]{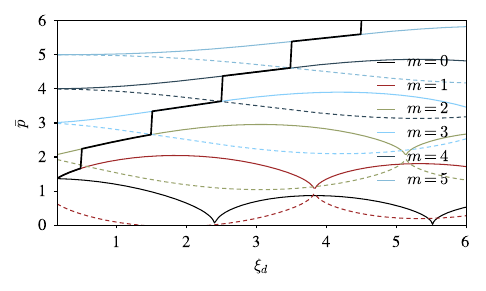}
    \caption{Estimate for the boundary of the chaotic region (thick black) as a function of the drive amplitude $\xi_d$, as derived from Chirikov's heuristic resonance overlap criterion. In thin lines, we plot $p_m^+$ (solid) and $p_m^-$ (dashed) for the first six separatrices in the upper half-plane $0\leq m \leq 5$ (the situation in the lower half plane is symmetric).}
    \label{fig:separatrices}
\end{figure}

Chirikov's heuristic overlap criterion states that chaotic regions arise whenever neighboring separatrices overlap. Thus, the upper bound of the chaotic layer is given by $\bar p = p^+_{\bar m}(0)$ where $\bar m \in \mathbb{N}$ is the largest integer such that for all $0\leq m <  \bar m$, resonance $m$ and $m+1$ overlap, \ie
\begin{align}
    p_m^+(0) > p_{m+1}^-(0).
\end{align}
Note that the situation of the separatrices is symmetric for negative values of $m$, which allows us to determine a theoretical value for the interval comprising the chaotic layer as $[-\bar{p},\bar{p}]$. We show the upper boundary of the chaotic layer as determined above in \cref{fig:separatrices}.

\subsection{Noise  Spectral Density}
\label{app:FGR}
In this section, we derive the spectral density of the diffusive process $p_t$ defined in \cref{eq:stochastic}. The invariant measure of the process is the uniform distribution over the interval $[-\bar{p},\bar{p}]$. We start with the calculation of the two-point correlation function under the assumption that $p_0$ is distributed according to the invariant measure, namely $p_0\sim \frac{1}{2\bar{p}} dy$. We have
\begin{align}
\begin{split}
  \mathbb E[p_0p_{\tau}] &= \int_{-\bar{p}}^{\bar{p}} \int_{-\bar{p}}^{\bar{p}}   \ x y d\mathbb P[p_0=y,p_{\tau}=x]  \\
  &= \int_{-\bar{p}}^{\bar{p}} \int_{-\bar{p}}^{\bar{p}} x y \frac{1}{2\bar{p}} dy d\mathbb P[p_{\tau}=x|p_0=y].
  \label{eq:C1}
\end{split}
\end{align}
In order to compute $d\mathbb P[p_{\tau}=x|p_0=y]$, we introduce $f_y(t,x)dx$, the probability density of the RBM at time $t>0$ starting from the initial distribution $\delta_y$. The Fokker-Planck equation for $f_y(t,x)$ reads
\begin{align}
    \frac{\partial }{\partial t} f(t,x) =\frac{D}{2} \frac{\partial^2 }{\partial x^2} f(t,x),
    \label{eq:FPeq}
\end{align}
with $D$ the diffusion rate \cref{eq:diffRate}. This equation is subject to the constraints implied by Neumann boundary conditions
\begin{align}
   \frac{\partial }{\partial x} f(t,x)\big |_{x=\bar{p}} = \frac{\partial }{\partial x} f(t,x)\big |_{x=-\bar{p}} =0.
   \label{eq:Neumann}
\end{align}
The next step is to calculate the Green's Function associated with this parabolic equation. To this aim, we need to diagonalize the operator $\frac{D}{2} \frac{\partial^2 }{\partial x^2}$ with Neumann boundary conditions. Its eigenfunctions read
\begin{align}
    \phi_0 = \frac{1}{\sqrt{2\bar{p}}}, \ \phi_n(x) = \frac{1}{\sqrt{\bar{p}}} \cos \left(\frac{n \pi}{2 \bar{p}} x-\frac{n \pi}{2}\right)
\end{align}
with associated eigenvalues
\begin{align}
    \lambda_n=-\left(\frac{n \pi}{2 \bar{p}}\right)^2 \frac{D}{2}.
    \label{eq:lambdan}
\end{align}
As a consequence, we have the following expression for the Green's function $f_y(t,x)$:
\begin{align}
    \begin{split}
       &f_y(t,x) = \sum_{n \geq 0} \langle \delta_y,\phi_n(x) \rangle  \phi_n(x)e^{\lambda_n t} \\
       &= \frac{1}{2\bar{p}}+\sum_{n \geq 1} \frac{e^{\lambda_n t}}{\bar{p}} \cos \left(\frac{n \pi}{2 \bar{p}} y-\frac{n \pi}{2}\right) \cos \left(\frac{n \pi}{2 \bar{p}} x-\frac{n \pi}{2}\right).
    \end{split}
\end{align}
Injecting into \cref{eq:C1} gives
\begin{align}
\begin{split}
  &\mathbb E[p_0p_{\tau}] = \frac{1}{2\bar{p}} \int_{-\bar{p}}^{\bar{p}} \int_{-\bar{p}}^{\bar{p}} x y   f_y(\tau,x)  dx dy   \\
  &= \frac{1}{2\bar{p}^2} \int_{-\bar{p}}^{\bar{p}} \int_{-\bar{p}}^{\bar{p}}  x y \\
   &\left[\frac{1}{2}+\sum_{n \geq 1} e^{\lambda_n \tau} \cos \left(\frac{n \pi x}{2 \bar{p}} -\frac{n \pi}{2}\right) \cos \left(\frac{n \pi y}{2 \bar{p}} -\frac{n \pi}{2}\right)  \right]dx dy,
\end{split}
\end{align}
where the first term, integrated over a symmetric domain, vanishes. Thus
\begin{align}
\begin{split}
  \mathbb E[p_0p_{\tau}] 
  = \sum_{n \geq 1} \frac{e^{\lambda_n \tau}}{2\bar{p}^2} \left[ \int_{-\bar{p}}^{\bar{p}}   x    \cos \left(\frac{n \pi x}{2 \bar{p}} -\frac{n \pi}{2}\right) dx   \right]^2.
\end{split}
\end{align}
The integral has a zero contribution for $n$ even, and a $4\left(\frac{2 \bar{p}}{n \pi}\right)^4$ contribution for $n$ odd, and hence
\begin{align}
\begin{split}
  \mathbb E[p_0p_{\tau}] 
  &= \frac{1}{2\bar{p}^2} \sum_{n \geq 1,odd} e^{\lambda_n \tau} 4\left(\frac{2 \bar{p}}{n \pi}\right)^4 \\
  &= \frac{32 \bar{p}^2}{\pi^4}  \sum_{n \geq 1,\text{odd}} \frac{1}{n^4}e^{\lambda_n \tau} .
\end{split}
\end{align}
Given that the sequence \(\left(\frac{1}{n^4}\right)_n\) decays rapidly and the rate at which \(e^{\lambda_n t} \to 0\) as \(t \to \infty\) increases with $n$, retaining only the first and second terms provides an excellent approximation.
 As a consequence, we get
\begin{align}
   \mathbb E[p_0p_{\tau}] \approx \frac{32 \bar{p}^2}{\pi^4} \left[ e^{\lambda_1 \tau}+\frac{1}{3^4} e^{\lambda_3 \tau}\right],
   \label{eq:corrf}
\end{align}
with $\lambda_n$ defined in \cref{eq:lambdan}.
Furthermore, the resulting spectral density reads
\begin{align}
    \begin{split}
        S_{pp}[\omega] &= \int_{-\infty}^{\infty} \mathbb E[p_0p_{|\tau|}] e^{-i\omega \tau} d\tau \\
        &= \frac{32 \bar{p}^2}{\pi^4} \left[\frac{2a}{\omega^2+a^2}+\frac{18a}{\omega^2+(9a)^2}\right]
        \label{eq:noiseSD}
    \end{split}
\end{align}
with $a=\frac{\pi^2 D}{8 \bar{p}^2}.$

\section{Spectral properties of the driven transmon}
\label{ap:DriTr}
\begin{figure}[t!]
    \centering
    \includegraphics[width=\linewidth]{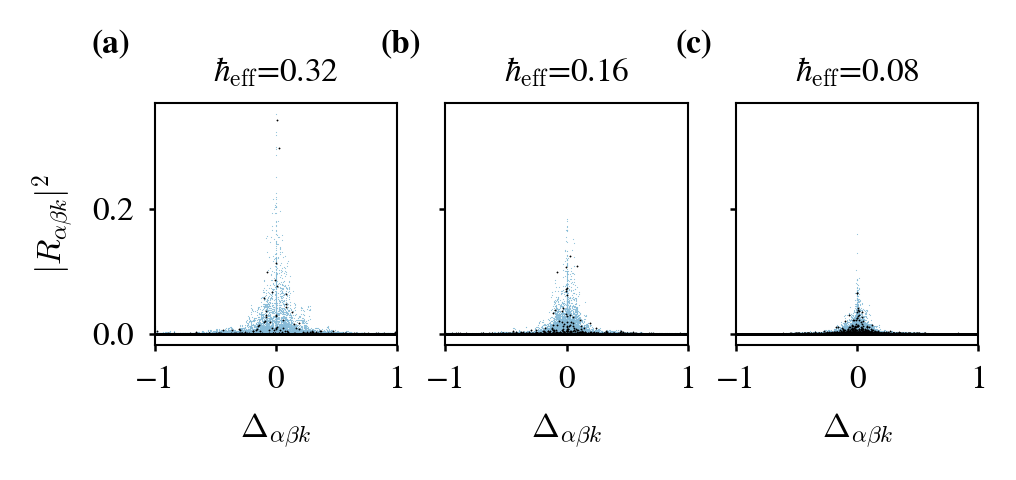}
    \caption{Weighted matrix element $R_{\alpha \beta k}$ of \cref{eq:Fcoeff2}, versus the transition frequency $\Delta_{\alpha \beta k}$ of the driven transmon at $\xi_d=1.5$ and $\lambda = 0.47$. From panel (a) to panel (c) $\hbar_{\text{eff}}$ (the number of transmon states) decreases (increases). To generate statistics we are varying the offset charge $n_g$ in the interval $[0,0.5]$, with black points corresponding to $n_g = 0$. The resulting distribution is peaked at zero frequency.}
    \label{fig:densityStates}
\end{figure}

In this section, we compute the Floquet spectrum of the driven transmon \cref{eq:Htr}, decoupled from the TLS, and argue that there exists an analogy between this spectrum and the one of an infinite bath. 
To make this analogy, consider the matrix element \cite{Grifoni1998DrivenQT}
\begin{align}
    P_{\alpha \beta k} = \frac{1}{T} \int_0^T dt \ e^{-i k  t} \langle \varphi_{\alpha}(t)|\hat p |\varphi_{\beta}(t) \rangle, 
    \label{eq:Fcoeff1}
\end{align}
corresponding to a transition of frequency $\Delta_{\alpha\beta k} = \eps_\alpha-\eps_\beta + k$ [recall that the drive frequency equals 1 in the units of \cref{eq:Schr}] between Floquet mode $\ket{\varphi_\beta(t)}$ and Floquet mode $\ket{\varphi_\alpha(t)}$ assisted by $k$ drive photons. In second-order perturbation theory, this matrix element participates in the Fermi's Golden rule transition rate of the transmon from state $\alpha$ to state $\beta$ \cite{bluemel_et_al_1991, Grifoni1998DrivenQT} under the action of a perturbation proportional to the operator $\hat{p}$. 

We assume that the transmon subsystem \cref{eq:Htr} is prepared in some initial chaotic state. To achieve this we prepare the transmon in its ground state and ensure that this state does not overlap with a regular island by  choosing an appropriate drive amplitude, as discussed in \cref{sec:Model}. 
We can now define a quantity consisting of the matrix element \cref{eq:Fcoeff1} weighed by the overlap of the ground state with each Floquet mode
\begin{align}
        R_{\alpha \beta k} = \langle \phi_{\beta}(0)|0 \rangle P_{\alpha \beta k}. 
    \label{eq:Fcoeff2}
\end{align}
\Cref{fig:densityStates} shows $|R_{\alpha \beta k}|^2$ versus the transition frequency $\Delta_{\alpha \beta k}$, for different values of $\hbar_{\text{eff}}$. We overlay plots for a sweep of the offset charge $n_g$ in $[0,0.5]$. 
As we decrease $\hbar_{\text{eff}}$, the spectrum, and hence the distribution of $|R_{\alpha \beta k}|^2$, becomes denser since the density of states  per energy window increases. In particular, we observe the symmetry of this distribution around zero frequency. This symmetry is responsible for similar excitation and relaxation rates, $\gamma_\uparrow$ and $\gamma_\downarrow$, as observed in~\cref{fig:gamma1}.

Furthermore, when a pair of Floquet modes is on resonance with $\Delta_{\alpha \beta k}=0$, the weighted matrix element $R_{\alpha \beta k}$ is significantly larger than at nonzero frequency, making zero-frequency noise dominant. This contribution is at the root of the qualitative discrepancy in the dephasing dynamics for the quantum  model on the one hand, and the semiclassical model with the RBM as noise source, on the other (\cref{fig:T2}).

\section{Details on numerical simulations}
\label{App:Num}
For our numerical diagonalization, we  use the basis formed by the transmon energy eigenstates, \ie eigenstates of the undriven Hamiltonian \cref{eq:Htr}. For example, in numerical simulations with $\hbar_{\text{eff}}=0.16$, we diagonalize a transmon Hamiltonian represented over $2D+1=401$ charge states (\ie integer-eigenvalued eigenstates of the Cooper pair number $\hat{n} = \sum_{n=-D}^D n \ket{n}\bra{n}$), then truncate the spectrum at $d=100$ transmon energy eigenstates, and express all coupling terms, \eg the coupling to the TLS in \cref{eq:Hfull}, in this new basis. The basis sizes above are chosen to ensure that, for all drive amplitudes considered here, the chaotic layer is fully captured in the numerical simulations, and the time-evolution plots are well converged. In addition to this, we check that, on the time-evolving wavefunction, the expectation value $\langle [ \hn, \cos\hphi] - i\sin \hphi \rangle$ is vanishing within a tolerance of $10^{-12}$. When we decrease $\hbar_{\text{eff}}$, we inversely proportionally consider more transmon states to accommodate for the increasing dimension of the transmon Hilbert space as the classical limit is approached.  The number of confined states inside the cosine potential, approximately given by $ 2 E_J/\sqrt{8 E_C E_J}= 2 \sqrt{\lambda} /\hbar_\text{eff}$, increases inversely proportionally to the decrease of $\hbar_{\text{eff}}$ at fixed plasma and drive frequencies. 

\end{appendix}

\bibliographystyle{apsrev4-2}
\bibliography{bibliography,circuit_qed}
\end{document}